# Multi-Mask Self-Supervised Learning for Physics-Guided Neural Networks in Highly Accelerated MRI

Burhaneddin Yaman[1,2], Hongyi Gu[1,2], Seyed Amir Hossein Hosseini[1,2], Omer Burak Demirel[1,2], Steen Moeller[2], Jutta Ellermann[2], Kâmil Uğurbil[2], Mehmet Akçakaya[1,2]

[1]Department of Electrical and Computer Engineering, and [2]Center for Magnetic Resonance Research, University of Minnesota, Minneapolis, MN

Word Count:

**Correspondence to:**

Mehmet Akçakaya, Ph. D.

University of Minnesota, 200 Union Street S.E., Minneapolis, MN, 55455

Phone: 612-625-1343; Fax: 612-625-4583

E-mail: akcakaya@umn.edu

**Funding:**

NIH, Grant numbers: R01HL153146, P41EB027061, U01EB025144; NSF, Grant number: CAREER CCF-1651825

# Abstract

Self-supervised learning has shown great promise due to its capability to train deep learning MRI reconstruction methods without fully-sampled data. Current self-supervised learning methods for physics-guided reconstruction networks split acquired undersampled data into two disjoint sets, where one is used for data consistency (DC) in the unrolled network and the other to define the training loss. In this study, we propose an improved self-supervised learning strategy that more efficiently uses the acquired data to train a physics-guided reconstruction network without a database of fully-sampled data. The proposed multi-mask self-supervised learning via data undersampling (SSDU) applies a hold-out masking operation on acquired measurements to split it into multiple pairs of disjoint sets for each training sample, while using one of these pairs for DC units and the other for defining loss, thereby more efficiently using the undersampled data. Multi-mask SSDU is applied on fully-sampled 3D knee and prospectively undersampled 3D brain MRI datasets, for various acceleration rates and patterns, and compared to CG-SENSE and single-mask SSDU DL-MRI, as well as supervised DL-MRI when fully-sampled data is available. Results on knee MRI show that the proposed multi-mask SSDU outperforms SSDU and performs closely with supervised DL-MRI. A clinical reader study further ranks the multi-mask SSDU higher than supervised DL-MRI in terms of SNR and aliasing artifacts. Results on brain MRI show that multi-mask SSDU achieves better reconstruction quality compared to SSDU. Reader study demonstrates that multi-mask SSDU at R=8 significantly improves reconstruction compared to single-mask SSDU at R=8, as well as CG-SENSE at R=2.

**Abbreviations**: CS, compressed sensing; DL, deep learning; DC, data consistency; SSDU, self-supervision via data undersampling; PG-DL, physics-guided deep learning; CG, conjugate gradient; RB, residual block; SNR, signal-to-noise ratio

# Introduction

Data acquisition is lengthy in many MRI exams, creating challenges for improving resolution and coverage, hence making accelerated MRI reconstruction an ongoing research topic. Parallel imaging (1-3) and compressed sensing (4-9) are two commonly used acceleration methods, with the former being the clinical gold standard for fast MRI, and the latter providing additional acceleration in a number of scenarios. However, acceleration rates remain limited as reconstructed images may suffer from noise amplification (10) or residual artifacts (11,12) in parallel imaging and compressed sensing, respectively. Recently, deep learning (DL) has emerged as an alternative for accelerated MRI due to its improved reconstruction quality compared to conventional approaches, especially at higher acceleration rates (13-18).

Among DL methods, physics-guided DL approaches, which incorporate the MRI encoding operator to solve a regularized inverse problem, have gained interest due to its robustness (18,19). Physics-guided DL approaches unroll an iterative process that alternates between data consistency (DC) and regularization for certain number of iterations. They are trained end-to-end, typically in a supervised manner by minimizing the difference between network output and a ground-truth reference obtained from fully-sampled data (14,20-22) . However, acquisition of fully-sampled data, especially on large patient populations, is either challenging or impossible in many practical scenarios (23-27).

As supervised training becomes inoperative in the absence of fully-sampled data, several methods have been proposed to train networks without fully-sampled data (28-33). Among these approaches, Self-supervision via Data Undersampling (SSDU) trains physics-guided neural networks by utilizing only the acquired sub-sampled measurements (31). In SSDU, the available measurements are split into two disjoint sets by a masking operation, which reduces the sensitivity to overfitting and is central for reliable performance. One of these sets is used in the DC units of the network, and the other is used to define the loss function in k-space. For moderately high acceleration rates, the networks trained using SSDU match the performance of those from supervised learning. While SSDU demonstrated that the splitting of acquired points into two sets was sufficient for training a neural network for reconstruction from undersampled data, a strategy that augments the use of the subsampled data to improve the performance is essential for higher acceleration rates.

In this study, we sought to improve the performance of SSDU with multiple masks. The proposed multi-mask SSDU splits acquired measurements into multiple pairs of disjoint sets for each training slice, while using one of these sets in a pair for DC units and the other for defining loss, similar to the original SSDU. The proposed multi-mask SSDU approach is applied on fully-sampled 3D knee MRI datasets from mridata.org (34), as well as a prospectively undersampled high-resolution 3D brain MRI dataset, and compared to parallel imaging, SSDU with a single mask (31), and supervised DL-MRI when fully-sampled data is available. Results show that the proposed multi-mask SSDU approach at high acceleration rates significantly improves upon SSDU and closely performs with supervised DL-MRI, while the reader studies indicate that the proposed multi-mask approach also outperforms supervised DL-MRI approach in terms of SNR improvement and aliasing artifact reduction.

## Methods

### Supervised Training of Physics-Guided DL-MRI Reconstruction

Let $\mathbf{y}_\Omega$ be the acquired subsampled measurements with $\Omega$ denoting the subsampling pattern and $\mathbf{x}$ the image to be recovered. The forward model for encoding is

$$\mathbf{y}_\Omega = \mathbf{E}_\Omega \mathbf{x} + \mathbf{n}, \qquad [1]$$

where $\mathbf{E}_\Omega: \mathbb{C}^M \to \mathbb{C}^P$ is the encoding operator including the coil sensitivities and a partial Fourier matrix sampling the locations specified by $\Omega$, and $\mathbf{n} \in \mathbb{C}^P$ is measurement noise. For sub-Nyquist sampling at high rates, the forward model may be ill-conditioned, necessitating the use of regularization, leading to an inverse problem for image reconstruction:

$$\arg\min_x \|\mathbf{y}_\Omega - \mathbf{E}_\Omega \mathbf{x}\|_2^2 + \mathcal{R}(x), \qquad [2]$$

where the first term represents DC and second term, $\mathcal{R}(\cdot)$ is the regularizer. Several approaches may be used to iteratively solve the above optimization problem (35). In this work, we use variable splitting via quadratic penalty method (21,31,35), which decouples DC and regularizer operations:

$$\mathbf{z}^{(i-1)} = \arg\min_z \mu \|\mathbf{x}^{(i-1)} - \mathbf{z}\|_2^2 + \mathcal{R}(z), \qquad [3]$$

$$\mathbf{x}^{(i)} = \arg\min_x \|\mathbf{y}_\Omega - \mathbf{E}_\Omega \mathbf{x}\|_2^2 + \mu \|\mathbf{x} - \mathbf{z}^{(i-1)}\|_2^2, \qquad [4]$$

where $\mu$ is the quadratic penalty parameter, $\mathbf{x}^{(i)}$ is the network output at iteration $i$ and $\mathbf{z}^{(i)}$ is an intermediate variable, and $\mathbf{x}^{(0)}$ is the initial image obtained from zero-filled under-sampled k-space data. In physics-guided DL, this iterative optimization is unrolled for a fixed number of iterations. Eq. [3] corresponds to a regularizer, which is implicitly solved by a neural network, whereas Eq. [4] has a closed form solution (31) that can be solved by gradient descent methods such as conjugate gradient (20).

In traditional DL-MRI approaches, training datasets contain pairs of undersampled k-space/ image and fully-sampled k-space/ground-truth image (14,20,21,36). Let $\mathbf{y}_{ref}^i$ be the fully-sampled reference k-space data for subject $i$, and $f(\mathbf{y}_\Omega^i, \mathbf{E}_\Omega^i; \boldsymbol{\theta})$ denote the output of the unrolled network that is parametrized by $\boldsymbol{\theta}$ for subsampled k-space data $\mathbf{y}_\Omega^i$ and corresponding encoding matrix $\mathbf{E}_\Omega^i$ of the same subject $i$. The supervised PG-DL training is performed by defining the loss function in image domain or k-space (31). Training can be performed by minimizing a k-space loss function as

$$\min_{\boldsymbol{\theta}} \frac{1}{N} \sum_{i=1}^{N} \mathcal{L}\left(\mathbf{y}_{ref}^i, \mathbf{E}_{full}^i \left(f(\mathbf{y}_\Omega^i, \mathbf{E}_\Omega^i; \boldsymbol{\theta})\right)\right), \qquad [5]$$

where $N$ is the number of fully-sampled training data in the database, $\mathbf{E}_{full}^i$ is the fully-sampled encoding operator that transforms network output to k-space, and $\mathcal{L}(.,.)$ is the loss between the fully-sampled and reconstructed k-spaces. The sampling locations, $\Omega$ may vary per subject in a more general setup, i.e. indexed by $i$. However, this was not included for simplicity of notation.

**Self-Supervision via Data Undersampling (SSDU)**

In order to enable training without fully-sampled datasets, SSDU has been proposed (31), where the acquired sub-sampled data indices, $\Omega$, from each scan are divided into two disjoint sets $\Theta$ and $\Lambda$. $\Theta$ is used in DC units in the unrolled network and $\Lambda$ is used to define the loss function, and the following self-supervised loss function is minimized

$$\min_{\boldsymbol{\theta}} \frac{1}{N} \sum_{i=1}^{N} \mathcal{L}\left(\mathbf{y}_\Lambda^i, \mathbf{E}_\Lambda^i \left(f(\mathbf{y}_\Theta^i, \mathbf{E}_\Theta^i; \boldsymbol{\theta})\right)\right). \qquad [6]$$

Unlike the supervised approach, only a subset of measurements, $\Theta$ are used as the input to the unrolled network. The network output is transformed to k-space, where the loss is calculated only at unseen k-space indices, Λ. After training is completed, testing is performed on unseen data using all available measurements Ω.

**Proposed Multi-mask SSDU**

SSDU reconstruction quality degrades at very high acceleration rates due to higher data scarcity, arising from the splitting into Θ and Λ. In order to tackle this issue, we propose a multi-mask SSDU approach, which retrospectively splits acquired indices Ω into disjoint sets Θ and Λ multiple times as shown in **Figure 1**. Formally, we split available measurements multiple times for each subject $i$ such that for each partition $\Omega = \Theta_j \cup \Lambda_j$, for j = 1,…, $K$ denoting the number of partitions for each scan. Similar to SSDU, each pair of sets in each scan were disjoint, i.e. $\Lambda_j = \Omega \backslash \Theta_j$ for j ∈ 1, …, $K$. In other words, Ω is retrieved by the union of each pair of $\Lambda_j$ and $\Theta_j$ for any j ∈ 1, …, $K$. Hence, the loss function to minimize during training becomes

$$\min_{\boldsymbol{\theta}} \frac{1}{N \cdot K} \sum_{i=1}^{N} \sum_{j=1}^{K} \mathcal{L}\left(\mathbf{y}_{\Lambda_j}^i, \mathbf{E}_{\Lambda_j}^i\left(f\left(y_{\Theta_j}^i, \mathbf{E}_{\Theta_j}^i; \boldsymbol{\theta}\right)\right)\right). \qquad [7]$$

The proposed multi-mask approach enables efficient use of available data by ensuring a higher fraction of low and high frequency components are utilized in the training and loss masks. Such utilization was inherently limited in the original SSDU approach, since each acquired k-space point was either used in training or loss masks only once.

**3D Imaging Datasets**

Fully-sampled 3D knee dataset were obtained from mridata.org (34), which were acquired with approval from the local institutional review board on a 3T GE Discovery MR 750 system with an

8-channel knee coil array using a fast spin-echo (FSE) sequence. Relevant imaging parameters were: FOV = 160×160×154 mm$^3$, resolution = 0.5×0.5×0.6 mm$^3$, matrix size = 320×320×256.

Brain imaging was performed on healthy subjects using a standard Siemens 3D-MPRAGE sequence at a 3T Siemens Magnetom Prisma (Siemens Healthcare, Erlangen, Germany) system using a 32-channel receiver head coil-array (31). The imaging protocols were approved by the local institutional review board, and written informed consent was obtained from all participants before each examination for this HIPAA-compliant study. Relevant imaging parameters were: FOV = 224×224×157 mm$^3$, resolution = 0.7×0.7×0.7 mm$^3$, matrix size = 320×320×224, prospective acceleration R = 2 (uniform in $k_y$) and ACS lines = 32 (31).

The 3D k-space datasets were inverse Fourier transformed along the read-out direction, and these slices were processed individually. The knee and brain datasets were retrospectively undersampled to R = 8 using a uniform sheared 2D undersampling pattern (37). Additionally, for the knee datasets, where a fully-sampled reference is available, further undersampling was performed at R = 8 using uniform 1D and 2D ($k_y$-$k_z$) random, and 1D and 2D and Poisson undersampling masks. The undersampling masks are provided in **Supporting Information Figure S1**. Finally, knee datasets were also undersampled to R = 12 using 2D random and Poisson undersampling masks. A 24×24 and 32×32 ACS region in the $k_y$-$k_z$ plane were kept fully-sampled for knee and brain datasets, respectively (31). The training sets for both knee and brain datasets consisted of 300 slices from 10 subjects, formed by taking 30 slices from each subject. For knee MRI, 2 different subjects with 200 slices were used for validation in multi-mask hyperparameter tuning, and 8 other different subjects were used for testing of the final method. For brain dataset, the testing was performed on

9 different subjects. The proposed multi-mask SSDU approach was compared with parallel imaging method, CG-SENSE; blind compressed sensing method (Blind-CS) (38); zero-shot learning approach deep image prior (DIP) (39); state-of-the art supervised deep learning method when applicable (20) and state-of-the-art self-supervised learning approach (SSDU) (31) as SSDU performs better than counterpart unsupervised learning techniques (**Supporting Information Figure S2**) (40).

## Choice of Multi-Mask Hyperparameters

There are several tunable hyperparameters in multi-mask SSDU, including the number of partitions, $K$ in Eq. [7], as well as the distribution and size of $\Lambda$ as in SSDU. A variable-density Gaussian distribution was used for $\Lambda$ in (31) for a single mask. In this study, we used a uniformly random distribution for the proposed approach, as the benefits of a variable density distribution diminish with multiple masks (**Supporting Information Figure S3**). In (31), the size of $\Lambda$ was optimized to $\rho = |\Lambda|/|\Omega| = 0.4$, which is also the optimal choice for the distribution considered here (**Supporting Information Figure S4**). After these two hyperparameters were set, the number of partitions of each scan, $K$ was varied among 3, 5, 6, 7, 8 and 10 to optimize the remaining distinct hyperparameter of the multi-mask SSDU.

## Network and Training Details

The iterative optimization problem in Eq. [3] and [4] was unrolled for $T=10$ iterations. Conjugate gradient descent was used in DC units of the unrolled network (20,31). The proximal operator corresponding to the solution of Eq. [3] employs the ResNet structure used in SSDU (31). It comprises input and output convolution layers and 15 residual blocks (RB) each containing two convolutional layers, where the first layer is followed by a rectified linear unit (ReLU) and the second layer is followed by a constant multiplication layer. All layers had a kernel size of 3×3, 64 channels. The unrolled network which shares parameters across the unrolled iterations had a total of 592,129 trainable parameters. As in SSDU, a ResNet structure as used for the regularizer in Eq. [3], where the network parameters were shared across the unrolled network (31). Coil sensitivity maps were generated from 24×24 center of k-space using ESPIRiT (41).

As a pre-processing step, maximum absolute value of the k-space for each slice in the datasets was normalized to 1 in all cases. The real and imaginary parts of the complex MRI dataset were concatenated as two channels prior to inputting into the network. Separate networks for knee and brain datasets were trained using the Adam optimizer with a learning rate of $5 \cdot 10^{-4}$, by minimizing a normalized $\ell_1$-$\ell_2$ loss function defined as $\mathcal{L}(\boldsymbol{u}, \boldsymbol{v}) = \frac{\|\boldsymbol{u}-\boldsymbol{v}\|_2}{\|\boldsymbol{u}\|_2} + \frac{\|\boldsymbol{u}-\boldsymbol{v}\|_1}{\|\boldsymbol{u}\|_1}$ (31) with a batch size of 1 over 100 epochs. All training was performed using Tensorflow in Python, and processed on a workstation with an NVIDIA Tesla V100 GPU with 32 GB memory. Implementation of this method will be provided online https://github.com/byaman14/SSDU.

**Image Evaluation**

Quantitative assessment of experimental results was performed using normalized mean square error (NMSE) and structural similarity index (SSIM) when fully-sampled data was available as reference. Moreover, qualitative assessment of the image quality from different reconstruction methods was performed by an experienced radiologist. For knee MRI, proposed multi-mask SSDU was compared with ground-truth obtained from fully-sampled data, SSDU and parallel imaging method CG-SENSE, all at R = 8 with 2D uniform undersampling. For brain MRI, the proposed multi-mask SSDU was compared with SSDU at R = 8. Additionally, CG-SENSE approach at the acquisition acceleration R = 2 was evaluated to serve as the clinical baseline. The reader, with 15 years of experience for musculoskeletal and neuro imaging, was blinded to the reconstruction method, which were shown in a randomized order to avoid bias except for the knowledge of the reference image in knee MRI dataset. Evaluations were based on a 4-point ordinal scale, adopted from (14) for blurring (1: no blurring, 2: mild blurring, 3: moderate blurring, 4: severe blurring), SNR (1: excellent, 2: good, 3: fair, 4: poor), aliasing artifacts(1: none, 2:mild, 3: moderate, 4: severe) and overall image quality (1: excellent, 2: good, 3: fair, 4: poor). Wilcoxon signed-rank test was used to evaluate the scores with a significance level of $P < 0.05$.

## Results

### Number of Partitions for Multi-Mask SSDU

**Figure 2** shows the effect of the proposed multi-mask self-supervised network training at R=8 using 2D uniform sheared undersampling with varying number of masks, $K$= 3, 5, 6, 7, 8, 10, as well as the ground-truth reference and the zerofilled undersampled data. Multi-mask SSDU approach suppresses residual artifacts as $K$ increases from 3 to 6. At $K = 7$, the visible residual artifacts are removed completely. When $K$ is further increased to 8 and 10, residual artifacts reappear. The quantitative assessment on validation dataset further confirms this qualitative assessment. The median and interquartile range of SSIM values on validation set were 0.8256 [0.7980, 0.8507], 0.8260 [0.8002, 0.8516], 0.8264 [0.8016, 0.8527], 0.8267 [0.8027, 0.8537], 0.8263 [0.8007, 0.8519], 0.8257 [0.7989, 0.8511], and NMSE values were 0.0138 [0.0121, 0.0158], 0.0135 [0.0119, 0.0158], 0.0135 [0.0119, 0.0157], 0.0134 [0.0118, 0.0156], 0.0135

[0.0119, 0.0158], 0.0137 [0.0121, 0.0159] using Gaussian selection for $K \in 3, 5, 6, 7, 8, 10$, respectively. Hence, $K = 7$ used for the remainder of the study.

**3D Imaging Datasets**

**Figure 3** depicts representative reference and reconstruction results of the 3D knee dataset using CG-SENSE, Blind-CS, DIP, supervised DL-MRI, SSDU and proposed multi-mask SSDU, as well as the difference images of these methods with respect to the reference for 2D uniform sheared R = 8 undersampling mask. Red arrows indicate that CG-SENSE suffers from highly-visible residual artifacts and noise amplification. Blind-CS and DIP severely suffers from the blurring artifacts. SSDU alleviates these artifacts substantially, though residual artifacts remain. The proposed multi-mask SSDU further removes these artifacts for both slices shown, while achieving similar reconstruction quality to supervised DL-MRI for the first slice, and further reducing the residual aliasing artifacts visible in the supervised DL-MRI approach for the second slice. Quantitative metrics and difference images in the figure further confirm that multi-mask SSDU outperforms SSDU, while performing similarly to supervised DL-MRI. Additional experimental results on 3D knee dataset using different undersampling patterns at R=8 and R=12 are provided in **Supporting Information Figures S5, S6 and S7**. In all these experiments, multi-mask SSDU visibly and quantitatively outperforms SSDU, further reducing the gap between self-supervised learning and supervised DL-MRI.

**Table 1** summarizes the median and interquartile ranges for NMSE and SSIM values metrics for different undersampling masks and acceleration rates across the whole knee MRI test datasets. In all cases, CG-SENSE reconstruction is significantly outperformed by all the DL approaches.

Among DL approaches, supervised DL-MRI outperforms self-supervised learning methods, while multi-mask SSDU quantitatively improves upon SSDU.

**Figure 4** demonstrates CG-SENSE reconstruction of a slice of the 3D-MPRAGE dataset at prospective acceleration R = 2, as well as CG-SENSE, Blind-CS, DIP, SSDU and the proposed multi-mask SSDU approach at retrospective acceleration R = 8 using 2D uniform sheared undersampling mask. Blind-CS and DIP reconstructions significantly suffer from blurring artifacts. SSDU at high acceleration R = 8 achieves similar reconstruction quality as CG-SENSE at acquisition acceleration R=2. Multi-mask SSDU further improves reconstruction quality by suppressing the noise evident in SSDU and CG-SENSE.

**Image evaluation scores**

**Figure 5a** and **b** summarize the reader study results for knee and brain datasets using 2D uniform sheared R = 8 undersampling mask, respectively. For knee MRI, proposed multi-mask SSDU was rated highest in terms of SNR, with a statistically significant improvement over all methods except supervised DL-MRI. For blurring, ground truth data was rated better than all methods. In terms of aliasing artifacts and overall image quality, the proposed multi-mask SSDU approach was rated best compared to other methods and the ground truth. In terms of these two evaluation criteria, all DL-MRI approaches and the reference showed similar statistical behavior, except SSDU was statistically worse than proposed multi-mask SSDU and supervised approach in terms of aliasing artifacts. A more comprehensive comparison also containing reader scores for CG-SENSE is presented in **Supporting Information Figure S8**.

For brain MRI, DL-MRI reconstructions trained using the proposed multi-mask SSDU and SSDU approach at acceleration rate of 8 performed similar with CG-SENSE at acquisition R = 2 in terms of SNR and blurring. However, in terms of aliasing artifacts, the proposed multi-mask SSDU significantly outperformed its counterparts. In terms of overall image quality, both SSDU methods at R = 8 showed statistically significant improvement over CG-SENSE at R = 2, while the proposed multi-mask SSDU achieved the best performance.

## Discussion

In this work, we extended our earlier work on self-supervision via data undersampling, which trains physics-guided neural network without fully-sampled data, to a multi-mask setting where multiple pairs of disjoint sets were used for each training slice in the dataset. Training of physics-guided DL-MRI reconstruction without ground-truth data remains an important topic, since acquisition of fully-sampled data is either impossible or challenging in a number of scenarios (23-27). Among multiple methods proposed for this goal (28-30,42,43), self-supervision directly uses the acquired data without relying on generative models or intermediate estimates. Our work makes several contributions to these approaches. The main contribution of the proposed multi-mask self-supervised learning approach is to use the available undersampled data more efficiently to enable physics-guided DL training, by retrospectively splitting these data into multiple 2-tuple of sets for the DC units during training and for defining loss. Another resulting contribution of the proposed multi-mask SSDU is an alternative data augmentation strategy for DL-MRI reconstruction, via the retrospective hold-out masking of the acquired measurements multiple times, with potential applications even beyond self-supervised learning (44). Finally, we applied the proposed multi-mask approach on knee and brain MRI datasets using different undersampling and acceleration rates, showing its improved reconstruction performance compared to single mask SSDU approach.

Specifically, the extensive experimental results using different subsampling patterns on retrospectively subsampled 3D knee dataset at R = 8 and R = 12 show that the proposed multi-mask SSDU consistently outperforms SSDU, while performing similarly with the supervised DL-MRI approach. Similarly, on prospectively subsampled brain MRI, multi-mask SSDU at R = 8 enhances the reconstruction quality of SSDU, while achieving lower noise level compared to SSDU at R = 8 and CG-SENSE at the acquisition R = 2.

As mentioned earlier, the proposed multi-mask SSDU approach can be interpreted as an alternative technique for data augmentation in DL-MRI reconstruction. With the proposed multi-mask data augmentation, self-supervised training was rated higher than supervised training in the reader study for knee imaging by a musculoskeletal expert reader in terms of noise and aliasing artifacts. Furthermore, Figure 3 showed example slices, where multi-mask self-supervised learning showed better performance in handling artifacts compared to supervised DL-MRI despite having lower SSIM and NMSE values. While these observations may seem surprising at first, it is consistent with recent studies that show quantitative metrics may not always align with the reconstruction performance (45). Additionally, there are recent studies that show self-supervised deep learning approaches outperforming its supervised *counterparts* in various applications (46,47). These and other studies suggest that supervised learning may preclude discovery, hence it may not generalize well on unseen data or may not be as robust as self-supervised learning techniques (48). Another interesting finding from the reader study on knee data was the worse scores given to the fully-sampled ground truth compared to DL-MRI methods. The expert reader noted the low SNR of the fully-sampled acquisition, due to the high acquisition resolution compared to conventional clinical scans, which was substantially improved visually using the inherent noise reduction of DL-MRI reconstruction.

While the proposed multi-mask SSDU approach enhances the SSDU performance, it also has a longer training time by a factor of $K$ compared to SSDU due to the increased size of the training dataset. Due to these lengthy training times, holdout cross-validation was used for the hyperparameter selection sub-study for optimizing $K$ instead of n-fold cross-validation. Furthermore, while the proposed multi-mask approach enables data augmentation, helping overcome data scarcity and enhance reconstruction quality, it also bears the risk of overfitting. In a broader context, it is understood that data augmentation can lead to massive datasets, but when this idea is applied to augment initially limited datasets, it may result in overfitting (49). This phenomenon was also observed in our study as the reconstruction quality does not monotonically improve with increasing $K$, and residual artifacts reappear for $K \geq 8$. The problem of choosing the optimal size of the post-augmented dataset, which corresponds to $K$ in our setup, remains an open problem in the broader machine learning community and this challenge has been highlighted in a recent survey on data augmentation as "*There is no consensus as to which ratio of original to final dataset size will result in the best performing model*" (49). Hence, while we have optimized $K$, this was done using the same experimental settings of (31). We note that the optimal value for $K$ may differ based on the selection of hyperparameters, such as the number of epochs or the learning rate. Nonetheless, our results readily show that multi-mask SSDU improves upon the single-mask SSDU in terms of quantitative metrics for any choice of $K>1$, while also suggesting that it is not advantageous to choose a very high value of $K$, both from a performance perspective, and from a practical viewpoint due to the increased training time.

In this study, we have split measurements into training and loss sets in a disjoint manner as SSDU has shown that overlaps between these sets lead to degraded reconstruction performance (31). A uniformly random selection of masks was used in the multi-mask SSDU. This was motivated by the issue that splitting $\Omega$ based on a Gaussian random selection leads to selecting mostly low-frequency components from scarce data, especially at high acceleration rates. With a Gaussian selection of $\Lambda$, a multi-mask approach still tends to select low-frequency components for each mask. **Supporting Information Figure S3** showed that using uniformly random selection in combination with multi-mask selection may circumvent this issue, since this will ensure both low and high frequency are contained in the loss masks of each scan.

Although we concentrated on random selection of masks, another special type of multi-mask SSDU may be based on using a cyclic selection that ensures all acquired measurements are used for both training and loss. **Supporting Information Figure S9** shows comparison between cyclic multi-mask and multi-mask SSDU approach, with multi-mask SSDU showing better reconstruction quality. Multi-mask SSDU does not impose any bounds on $K$, allowing $K \times |\Lambda| \geq |\Omega|$, while cyclic multi-mask SSDU strictly imposes $K \times |\Lambda| = |\Omega|$. Hence, although cyclic multi-mask SSDU ensures every point in $\Omega$ is eventually used in both DC units and for defining the k-space loss, it inherently limits the number of masks for training, which in turn hinders the amount of improvement in reconstruction quality.

The multi-masking approach proposed in this work may also be adapted to the supervised learning setting by using multiple random $\Theta_j$ in the DC units of the unrolled network, while calculating the loss on the fully-sampled k-space. This extension to supervised training, which introduces an additional degree of randomness to the training process, was recently shown to improve performance over conventional supervised DL-MRI approach (50). However, we note that since this is a new extension arising from this work, comparisons in this study were made to the conventional DL-MRI approach that is used in the literature (20).

The proposed multi-mask SSDU approach also shares similarities with bootstrap aggregation mechanisms. In bootstrap approaches, multiple sub-datasets are generated by randomly sampling from a main dataset. The final prediction is performed by averaging outputs from each of these sub-dataset to reduce the variance among trained models. However, in multi-mask SSDU, each sample in the main dataset is sub-sampled multiple times by retrospectively splitting its measurements into disjoint sets. As a result, an aggregated large dataset which contains the measurements of each sample in the main dataset multiple times is obtained and used for training. Unlike bootstrapping approaches, the proposed approach performs final prediction by directly using the model trained on the aggregated large dataset.

The study has limitations. In the proposed multi-mask SSDU, we optimized the hyperparameters $\rho$ and $K$ independently for two main reasons: 1) If the joint optimization led to a different $\rho$ value, then this would create a confounding variable for the direct comparison to the single-mask scenario, 2) Optimizing over the two hyperparameters jointly leads to a large number of trainings for marginal gain. Such large number of trainings, which does not show a substantial implication in terms of the perspective of the study may also come at an environmental cost, as training of DL models have been shown to lead to considerable amount of carbon emissions (51). Thus, in the study, we concentrated on the individual optimization of the *K* parameter for the fixed ρ value that works best for single-mask SSDU (31), as this provides a more fair comparison. Additionally, this study focused on methodological development to improve the performance on single-mask SSDU without a specific application that may leverage large stores of existing undersampled data. With the methodology in place, such applications are being pursued in subsequent studies, both in LGE cardiac and brain MRI applications (52-55). Moreover, the proposed multi-mask SSDU may also be synergistically combined with dynamic cardiac MRI applications, in which spatio-temporal information can be leveraged to boost the performance at high acceleration rates (56-58).

More recently, zero-shot learning approaches have gained interest as they enable training using only a single slice (59-61). Thus, zero-shot learning approaches do not require any external dataset for training unlike database deep learning approaches, such as SSDU and proposed multi-mask SSDU. Deep image prior is the pioneer zero-shot learning approach, but it requires a heuristic early stopping criterion. Recent works on zero-shot learning have built on the multi-mask concept of the proposed approach to develop a rigorous stopping criterion (57). While zero-shot learning enables training from a single slice, the reconstruction times are much longer compared to database learning approaches.

## Conclusion

The proposed multi-mask SSDU approach enables training of physics-guided neural networks without fully-sampled data, while significantly outperforming single-mask SSDU at high acceleration rates through the efficient use of the acquired undersampled data with multiple masks.

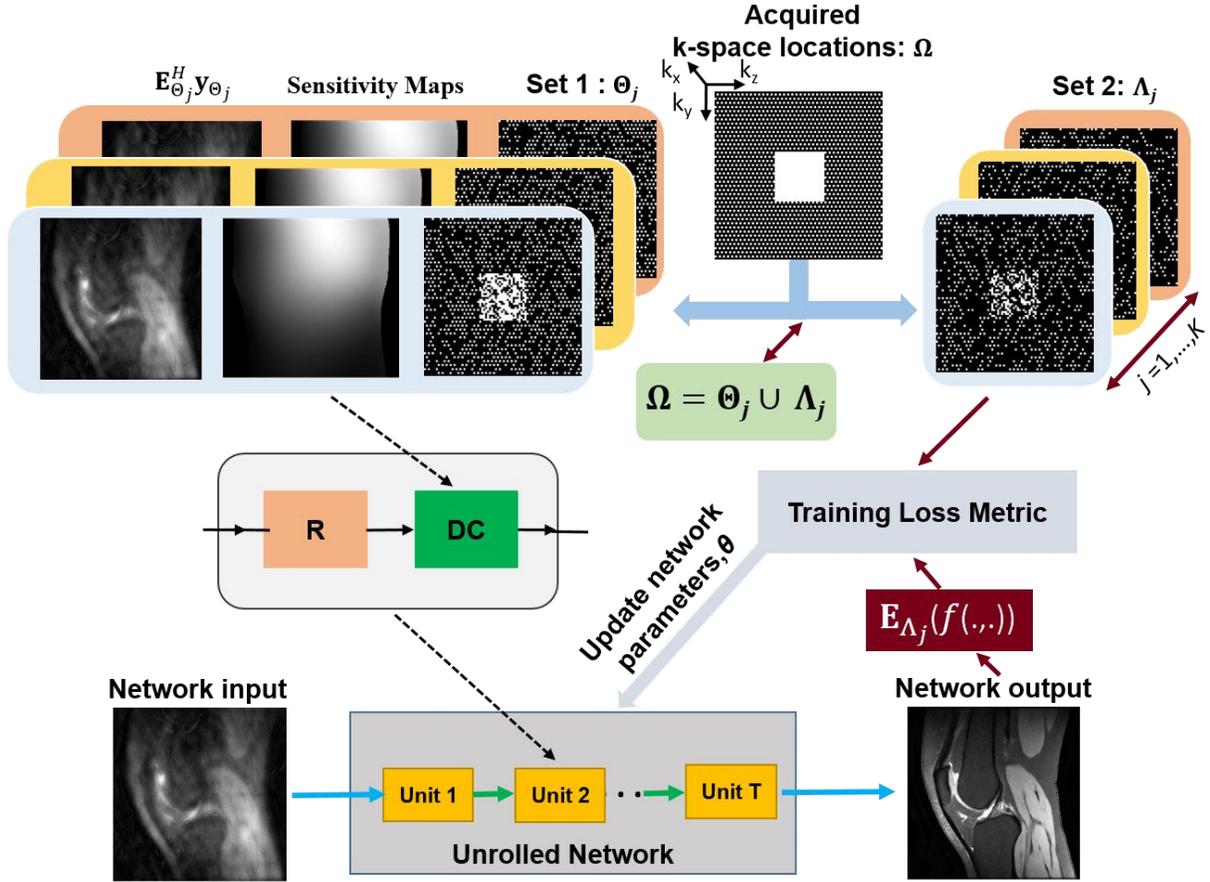

**Figure 1.** The multi-mask self-supervised learning scheme to train physics-guided deep learning without fully-sampled data. The acquired sub-sampled k-space measurements for each scan, $\Omega$, are split into multiple disjoint sets, $\Theta_j$ and $\Lambda_j$ in which $\Omega = \Theta_j \cup \Lambda_j$, for $j \in 1, ..., K$. The first set of indices, $\Theta_j$, is used in the data consistency unit of the unrolled network, while the latter set, $\Lambda_j$ is used to define the loss function for training. During training, the output of the network is transformed to k-space, and the available subset of measurements at $\Lambda_j$ are compared with the corresponding reconstructed k-space values. Based on this training loss, the network parameters are subsequently updated.

**Figure 2.** A representative test slice showing the reconstruction results for different number of partitions *K* using 2D uniform sheared R = 8 undersampling mask. Red arrows mark residual artifacts for *K*≤6 and *K*≥8. These artifacts are suppressed at *K*=7, which is used for the remainder of the study.

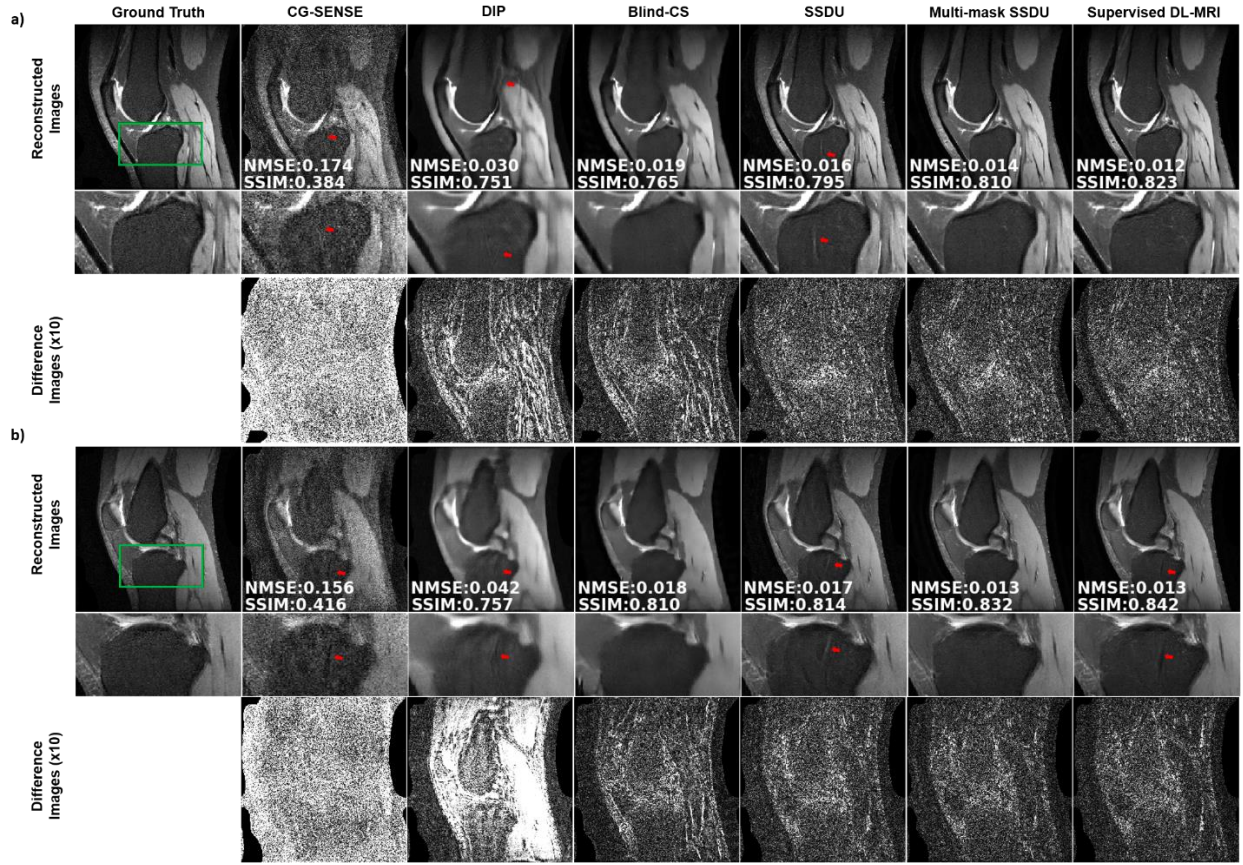

**Figure 3.** a) and b) Representative test slices from 3D FSE knee MRI dataset showing the reconstruction results for proposed multi-mask self-supervised DL-MRI (multi-mask SSDU), self-supervised DL-MRI (SSDU), supervised DL-MRI, DIP, Blind-CS, and CG-SENSE approaches for retrospective 2D uniform undersampling R = 8, as well as the error images with respect to the fully-sampled reference. CG-SENSE suffers from substantial residual artifacts that are shown with red arrows for both slices. Blind-CS and DIP severely suffers from the blurring artifacts. DL-MRI with SSDU learning suppresses a large portion of these artifacts, but still exhibits visible residual artifacts in both scenarios. Proposed multi-mask SSDU successfully suppresses these artifacts further for both slices, in a) closely matches the performance of supervised DL-MRI and in b) reduces residual aliasing further compared to supervised DL-MRI.

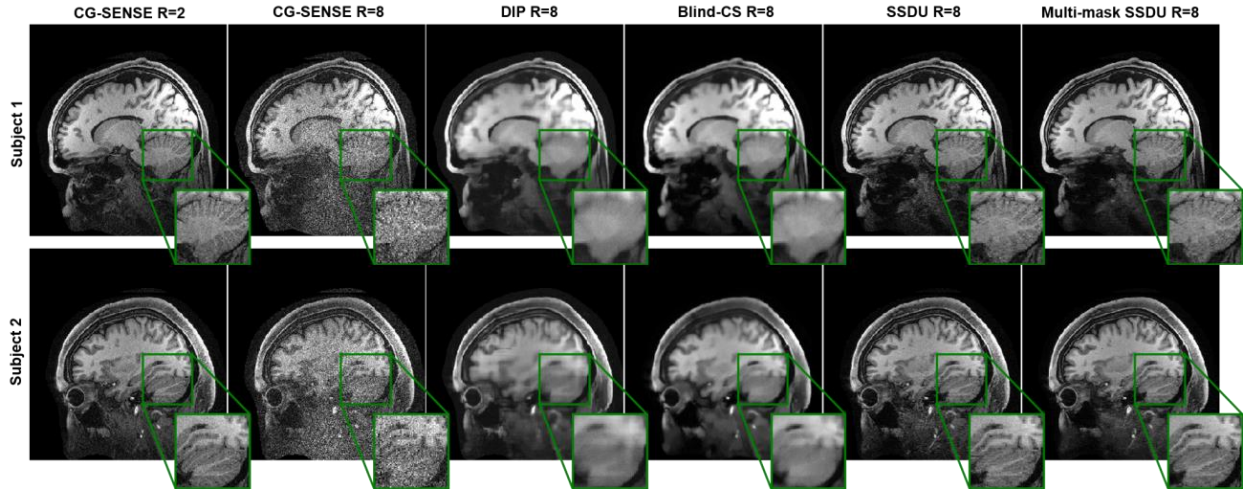

**Figure 4.** Reconstruction results from prospectively 2-fold equispaced undersampled brain MRI. SSDU, multi-mask SSDU, DIP, Blind-CS and CG-SENSE are applied at further retrospective acceleration rates of 8 with equispaced sheared $k_y$-$k_z$ undersampling patterns, while CG-SENSE is also used at the acquisition rate of 2, which serves as the clinical baseline. CG-SENSE suffers from visibly higher noise amplification at R = 8. Blind-CS and DIP reconstructions significantly suffer from blurring artifacts. SSDU DL-MRI performs successful reconstruction at R = 8, while achieving similar image quality to CG-SENSE at R = 2. The proposed multi-mask SSDU DL-MRI further enhances the SSDU DL-MRI performance by achieving lower noise level in reconstruction results.

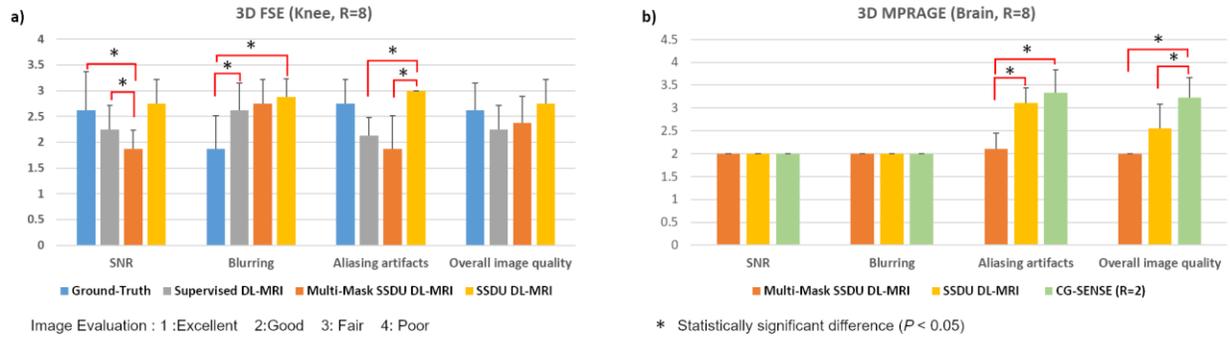

**Figure 5.** a) Reader study for knee MRI for reconstructions using uniform 2D undersampling masks at R=8. Bar-plots show average reader scores and their standard deviation across the test subjects. Statistical testing was performed by one-sided Wilcoxon single-rank test, with * showing significant statistical difference with $P <0.05$. In terms of SNR, the proposed multi-mask SSDU was rated highest, and statistically better than all approaches except supervised DL-MRI. For blurring, ground truth data was rated statistically better than all methods except the proposed multi-mask SSDU. In terms of aliasing artifacts and overall image quality, the proposed multi-mask SSDU approach was rated best compared to other methods and ground truth. In terms of these two evaluation criteria, all DL-MRI approaches and the reference showed similar statistical behavior, except SSDU was statistically worse than proposed multi-mask SSDU and supervised approach in terms of aliasing artifacts. b) Reader study for brain MRI. CG-SENSE at R = 2, and proposed multi-mask SSDU and SSDU at R = 8 using uniform 2D undersampling mask were in good agreement in terms of SNR and blurring. In terms of aliasing artifacts and overall image quality, the proposed multi-mask SSDU approach received the best scores, while CG-SENSE at R = 2 was rated lowest and showed significant statistical difference with proposed multi-mask SSDU in terms of both evaluation criteria and SSDU in terms of overall image quality. The proposed multi-mask SSDU was also rated statistically better than SSDU in terms of aliasing artifacts.

|  |  | CG-SENSE | DIP | Blind-CS | Supervised DL-MRI | SSDU | Multi-mask SSDU |
|---|---|---|---|---|---|---|---|
| Uniform 1D R=8 | NMSE | 0.0994 [0.0949, 0.1419] | 0.0457 [0.0316, 0.0627] | 0.0192 [0.0172, 0.2274] | 0.0122 [0.0108, 0.0139] | 0.0166 [0.0147, 0.0191] | 0.0140 [0.0124, 0.0161] |
|  | SSIM | 0.4698 [0.4193, 0.5353] | 0.7224 [0.6819, 0.7435] | 0.7945 [0.7648, 0.8332] | 0.8505 [0.8306, 0.8751] | 0.8211 [0.7947, 0.8527] | 0.8315 [0.8084, 0.8600] |
| Uniform 2D R=8 | NMSE | 0.1475 [0.1291, 0.1779] | 0.0438 [0.0297, 0.0622] | 0.0187 [0.0168, 0.0218] | 0.0124 [0.0112, 0.0143] | 0.0164 [0.0148, 0.0189] | 0.0135 [0.0123, 0.0155] |
|  | SSIM | 0.4411 [0.3797, 0.4976] | 0.7124 [0.6704, 0.7452] | 0.7884 [0.7526, 0.8126] | 0.8421 [0.8201, 0.8662] | 0.8150 [0.7877, 0.8426] | 0.8298 [0.8067, 0.8560] |
| Random 1D R=8 | NMSE | 0.0994 [0.0805, 0.1236] | 0.0392 [0.0216, 0.0524] | 0.0179 [0.0162, 0.0205] | 0.0121 [0.0107, 0.0139] | 0.0156 [0.0135, 0.0177] | 0.0137 [0.0121, 0.0155] |
|  | SSIM | 0.4886 [0.4305, 0.5571] | 0.7326 [0.7005, 0.7715] | 0.8032 [0.7789, 0.8407] | 0.8524 [0.8314, 0.8756] | 0.8328 [0.8089, 0.8615] | 0.8367 [0.8144, 0.8633] |
| Random 2D R=8 | NMSE | 0.1473 [0.1301, 0.1759] | 0.0412 [0.0256, 0.0567] | 0.0197 [0.0178, 0.0224] | 0.0130 [0.0117, 0.0149] | 0.0173 [0.0155, 0.0199] | 0.0145 [0.0131, 0.0165] |
|  | SSIM | 0.4239 [0.3631, 0.4766] | 0.7286 [0.6957, 0.7592] | 0.7914 [0.7648, 0.8229] | 0.8379 [0.8164, 0.8637] | 0.8123 [0.7853, 0.8417] | 0.8224 [0.8002, 0.8509] |
| Poisson R=8 | NMSE | 0.1035 [0.0937, 0.1206] | 0.0358 [0.0182, 0.0497] | 0.0158 [0.0134, 0.0178] | 0.0101 [0.0091, 0.0112] | 0.0131 [0.0118, 0.0149] | 0.0108 [0.0098, 0.0121] |
|  | SSIM | 0.4885 [0.4394, 0.5397] | 0.7325 [0.6914, 0.7629] | 0.8075 [0.7800, 0.8370] | 0.8554 [0.8365, 0.8793] | 0.8312 [0.8066, 0.8585] | 0.8421 [0.8212, 0.8679] |
| Random 2D R=12 | NMSE | 0.1331 [0.1207, 0.1556] | 0.0542 [0.0386, 0.0785] | 0.0247 [0.0218, 0.0287] | 0.0157 [0.0141, 0.0179] | 0.0221 [0.0198, 0.0254] | 0.0185 [0.0167, 0.0208] |
|  | SSIM | 0.4325 [0.3756, 0.4796] | 0.6834 [0.6576, 0.7182] | 0.7614 [0.7264, 0.7916] | 0.8148 [0.7916, 0.8431] | 0.7809 [0.7517, 0.8151] | 0.7982 [0.7722, 0.8288] |
| Poisson R=12 | NMSE | 0.0876 [0.0795, 0.1012] | 0.0491 [0.0342, 0.0673] | 0.0173 [0.0151, 0.0192] | 0.0119 [0.0107, 0.0133] | 0.0151 [0.0136, 0.0169] | 0.0129 [0.0117, 0.0145] |
|  | SSIM | 0.5119 [0.4638, 0.5597] | 0.7014 [0.6632, 0.7376] | 0.7753 [0.7385, 0.8164] | 0.8362 [0.8156, 0.8625] | 0.8133 [0.7862, 0.8442] | 0.8326 [0.8098, 0.8609] |

**Table 1.** The median and interquartile ranges for NMSE and SSIM metrics for different undersampling masks and acceleration rates. Note that due to the different size of the ACS data, 1D masks correspond to an effective acceleration rate of 5.2, while the 2D masks yield an effective acceleration rate of 7.7.

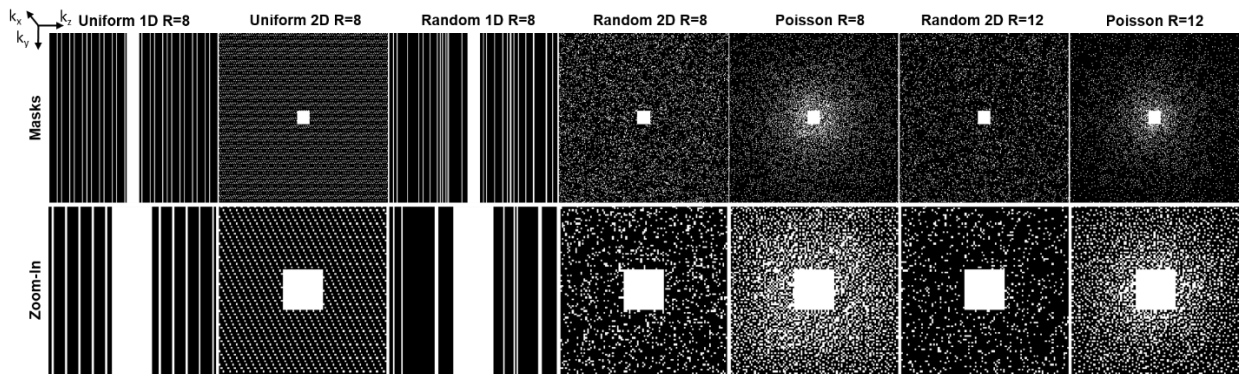

**Supporting Information Figure S1.** Undersampling masks used in the study. Note that due to the different size of the ACS data, 1D masks correspond to an effective acceleration rate of 5.2, while the 2D masks yield an effective acceleration rate of 7.7.

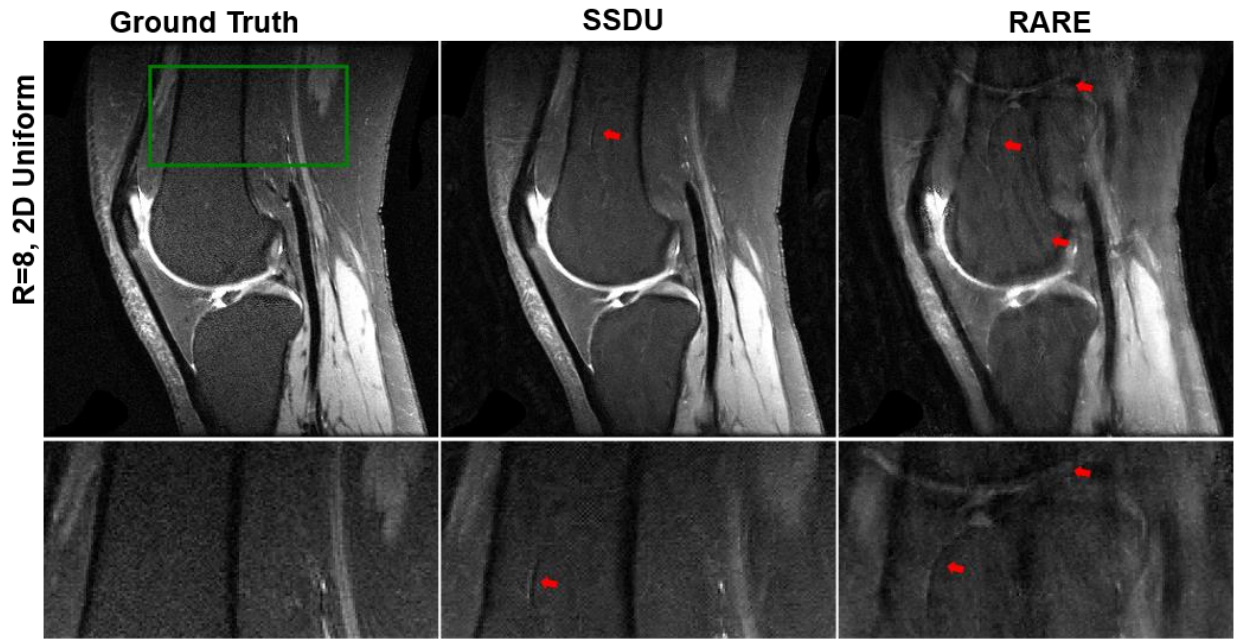

**Supporting Information Figure S2.** Reconstruction results from SSDU and RARE at R=8 using 2D uniform undersampling pattern. RARE suffers from major artifacts shown with red arrows, while SSDU achieves an improved reconstruction quality.

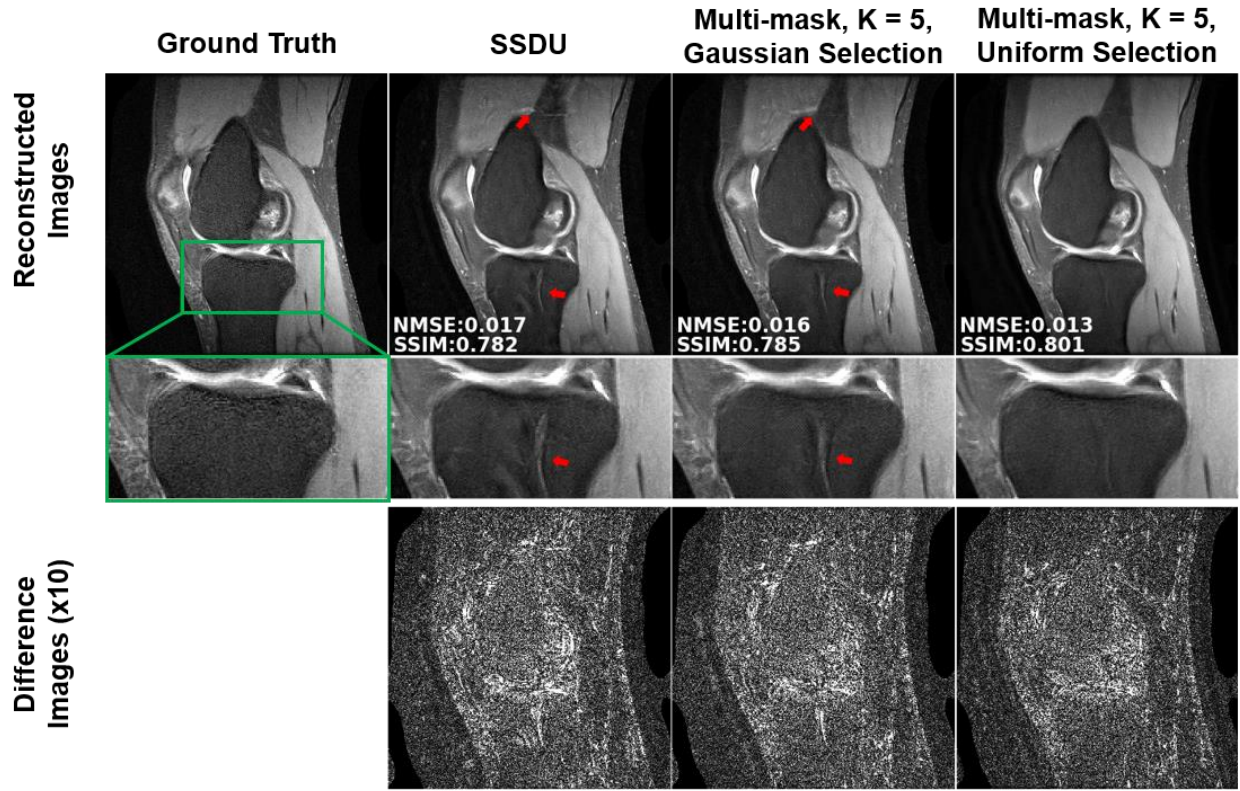

**Supporting Information Figure S3.** Reconstruction results from SSDU, and multi-mask SSDU at R=8 using 2D uniform undersampling mask with uniform random selection and variable-density Gaussian selection for K = 5 and ρ = 0.4. Multi-mask SSDU with Gaussian random selection fails to remove the artifacts apparent in SSDU, whereas multi-mask SSDU with uniformly random selection significantly suppresses these artifacts. Difference images show that multi-mask SSDU with uniformly random selection shows fewer residual artifacts compared to its multi-mask Gaussian counterpart. The median and interquartile range of SSIM values across the validation dataset were 0.7974 [0.7723, 0.8293], 0.8009 [0.7789, 0.8313], 0.8260 [0.8002, 0.8516], and NMSE values were 0.0166 [0.0142, 0.0202], 0.0159 [0.0139, 0.0191], 0.0135 [0.0119, 0.0157] for SSDU, multi-mask SSDU with Gaussian selection and uniformly random selection, respectively.

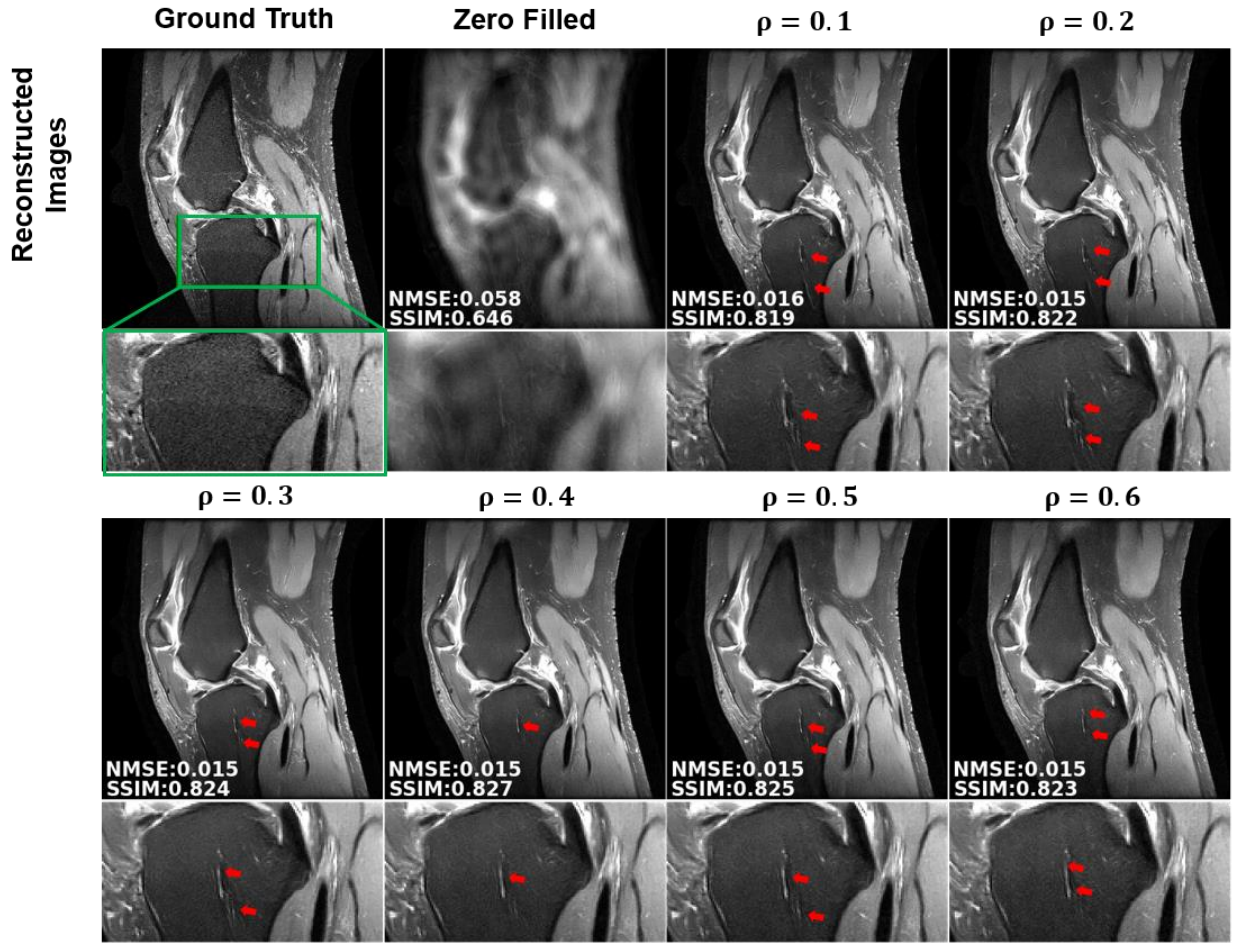

**Supporting Information Figure S4.** Reconstruction results from SSDU with uniform random selection of Λ for ρ ∈ {0.1, 0.2, 0.3, 0.4, 0.5, 0.6} using 2D uniform undersampling mask at R=8. SSDU reconstructions suffers from residual artifacts for low ρ values of 0.1, 0.2 and 0.3. The best reconstruction quality is achieved at ρ = 0.4. Residual artifacts start to reappear after ρ = 0.5, becoming more pronounced as ρ increases. The quantitative assessment from hold-out validation set align with these qualitative assessments. The median and interquartile range of SSIM values were 0.8166 [0.7875, 0.8408], 0.8208 [0.7928, 0.8451], 0.8230 [0.7967, 0.8486], 0.8236 [0.7964, 0.8494], 0.8229 [0.7960, 0.8499], 0.8192 [0.7937, 0.8473], and NMSE values were 0.0149 [0.0136, 0.0175], 0.0143 [0.0128, 0.0167], 0.0141 [0.0123, 0.0163], 0.0140 [0.0122, 0.0161], 0.0145 [ 0.0125, 0.0168], 0.0145 [0.0127, 0.0169] using uniformly random selection for ρ ∈ {0.1, 0.2, 0.3, 0.4, 0.5, 0.6}, respectively.

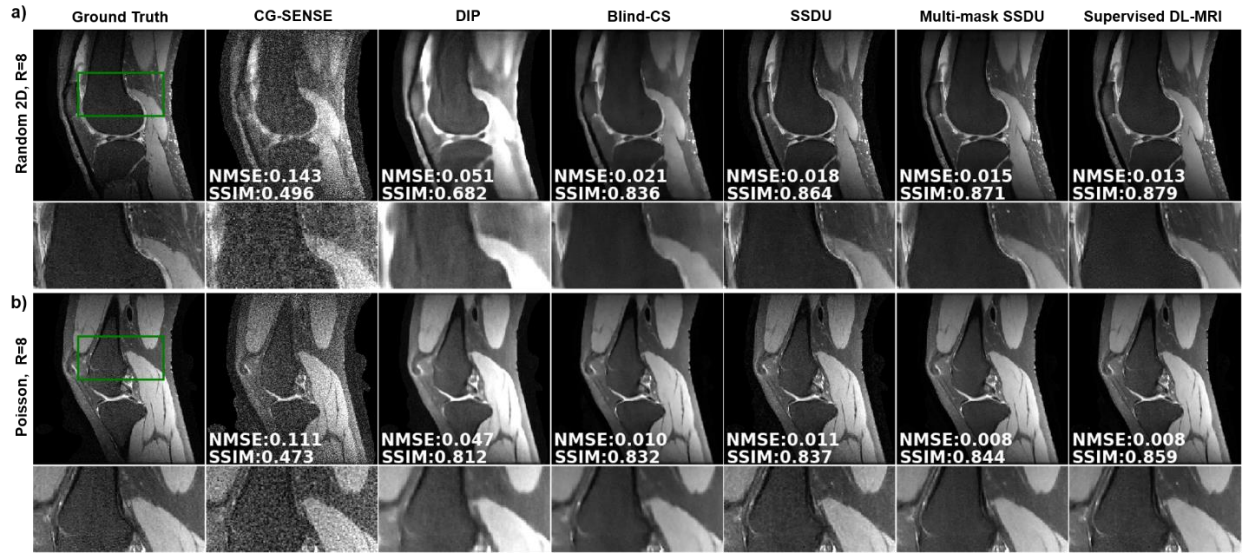

**Supporting Information Figure S5.** Reconstruction results using 2D a) random and b) Poisson undersampling masks at R = 8. CG-SENSE suffers from noise and incoherent residual artifacts. Blind-CS and DIP suffers from blurring artifacts. All other DL approaches achieve artifact-free and improved reconstruction quality.

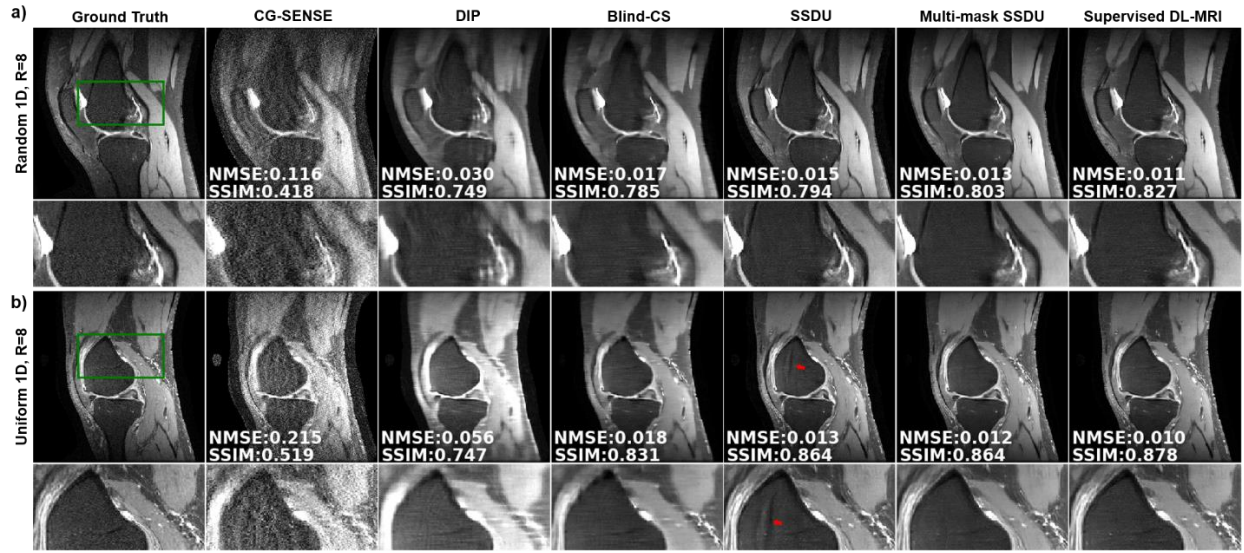

**Supporting Information Figure S6.** Reconstruction results at R = 8 using 1D a) random and b) uniform undersampling masks. CG-SENSE suffers from noise and residual artifacts for both of these undersampling masks. Blind-CS and DIP severely suffers from blurring artifacts. All other DL reconstructions achieve artifact-free reconstruction with random undersampling. In uniform undersampling, SSDU suffers from residual artifacts shown with red arrows, whereas multi-mask SSDU improves upon SSDU and achieve similar reconstruction quality with supervised DL-MRI.

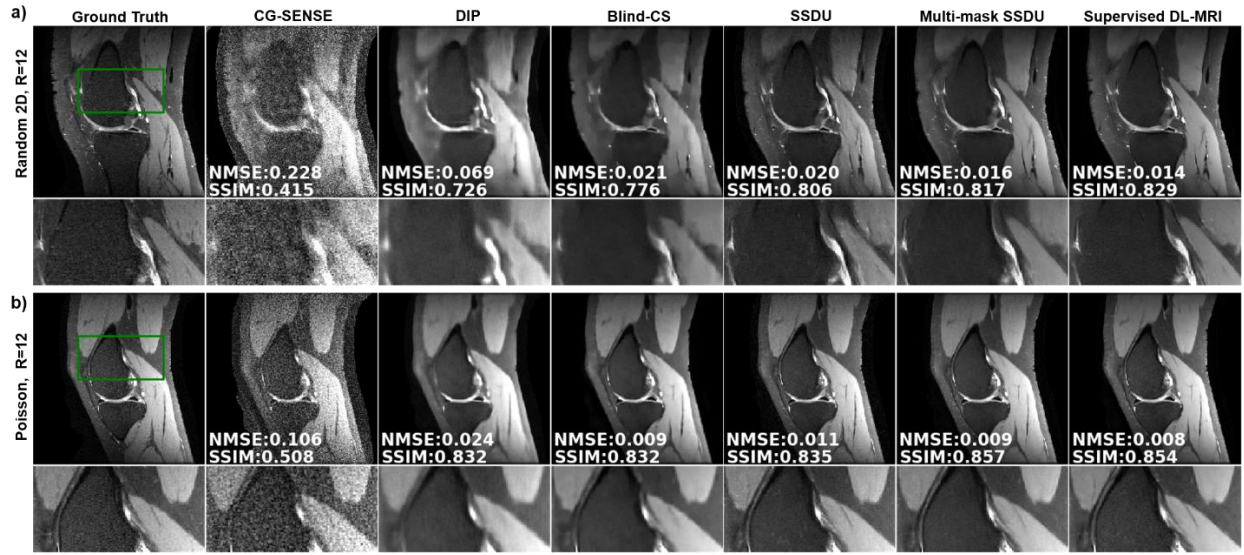

**Supporting Information Figure S7.** Reconstruction results at R = 12 using 2D a) random and b) Poisson undersampling masks. CG-SENSE suffers from noise and incoherent residual artifacts. Blind-CS and DIP suffers from visible blurring artifacts. All other DL approaches achieve artifact-free and improved reconstruction quality.

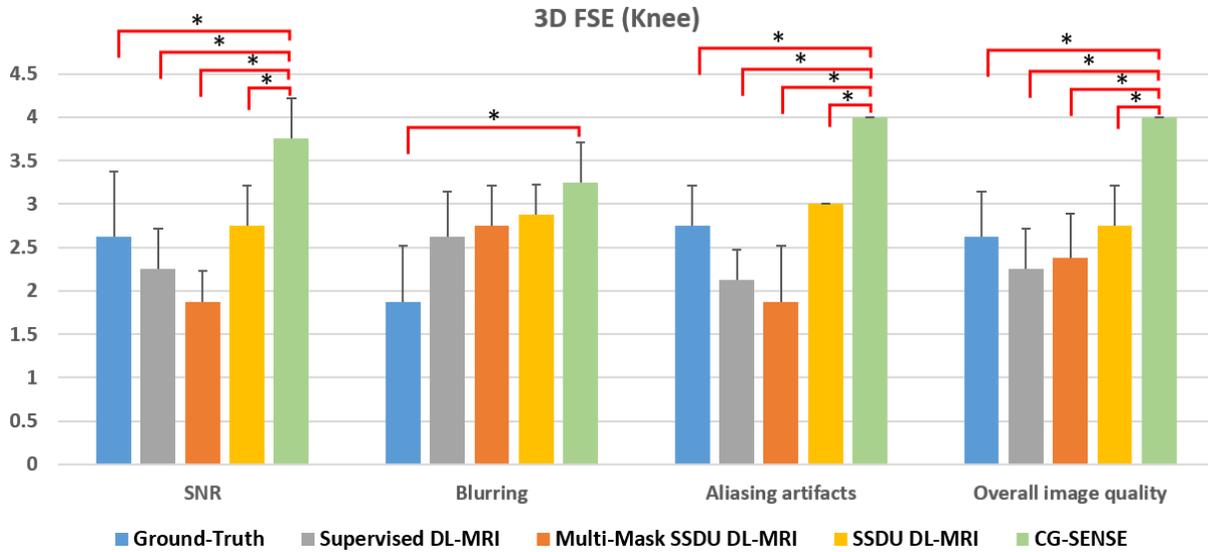

**Supporting Information Figure S8.** The image reading results from the clinical reader study for the 3D FSE knee dataset using 2D uniform sheared R = 8 undersampling mask. CG-SENSE was consistently rated lowest in terms of all evaluation criteria. CG-SENSE was significantly worse than all other methods and ground truth in terms of SNR, aliasing artifacts and overall image quality. For blurring, it was only statistically different than the ground truth.

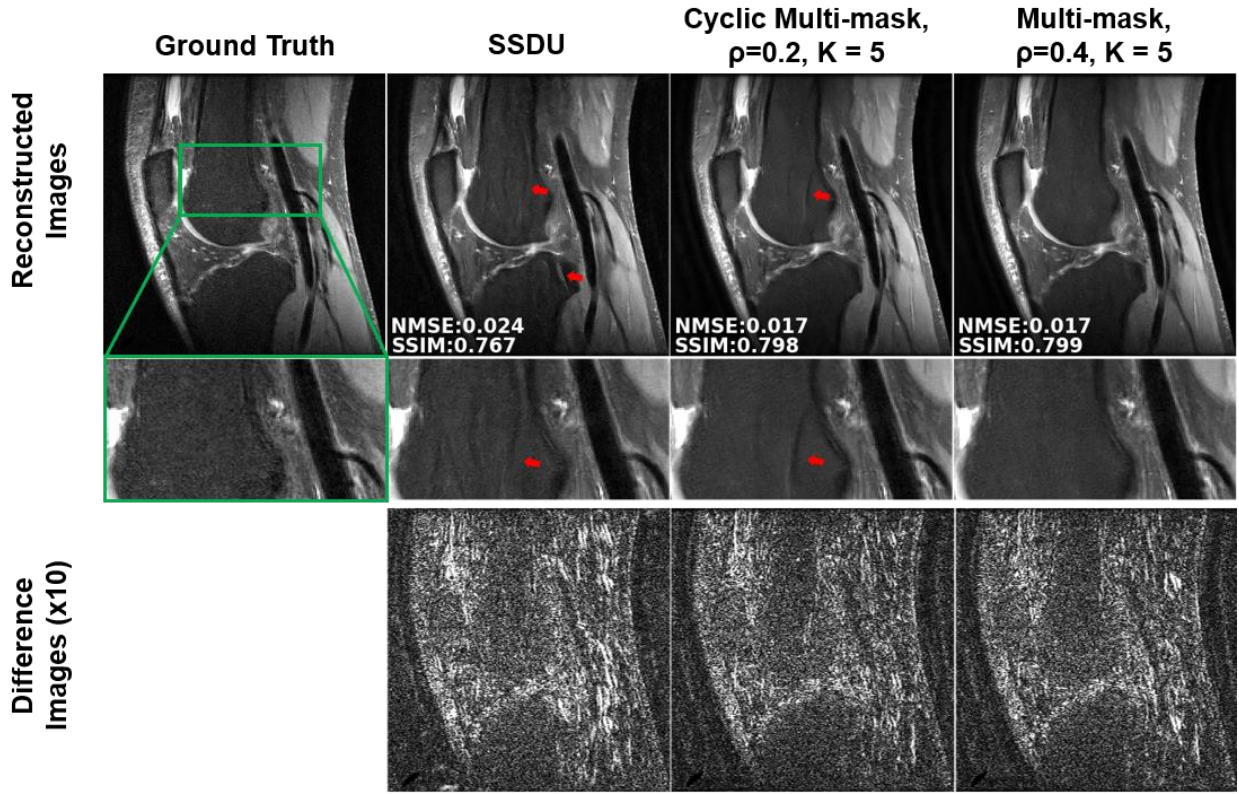

**Supporting Information Figure S9.** Reconstruction results from SSDU, multi-mask SSDU and multi-mask cyclic SSDU for K = 5 using 2D uniform undersampling masks at R=8. In multi-mask SSDU $\rho$ = 0.4 for K = 5, whereas multi-mask cyclic SSDU approach enforces $\rho$ to be 0.2 for K = 5. Multi-mask SSDU successfully removes artifacts in SSDU, whereas multi-mask cyclic SSDU suffers from residual artifacts. Difference images further confirm these observations. In this setting, the median and interquartile range of SSIM values across the validation dataset were 0.7974 [0.7723, 0.8293], 0.8249[0.7968, 0.8497], 0.8260 [0.8002, 0.8516], and NMSE values were, 0.0166 [0.0142, 0.0202], 0.0137 [0.0121, 0.0161], 0.0135 [0.0119, 0.0157] for SSDU, multi-mask cyclic SSDU and multi-mask SSDU, respectively.